\documentclass[nofootinbib,a4paper,fleqn,aps,prd,10pt,superscriptaddress,reprint,showkeys,showpacs]{revtex4-1}

\usepackage{times}
\usepackage{amsfonts}
\usepackage{amsmath}
\usepackage{amssymb}
\usepackage{natbib}
\usepackage{graphicx}
\usepackage{enumitem}
\usepackage{xcolor}
\usepackage[colorlinks=true,linkcolor=blue]{hyperref}
\usepackage[outdir=./]{epstopdf}
\usepackage[normalem]{ulem}

\def\aap{A\&A}

\def\apj{ApJ}
\def\apjl{ApJL}
\def\apjs{ApJS}
\def\apss{Astrophysics and Space Science}
\def\jcap{JCAP}
\def\mnras{MNRAS}

\def\jcap{JCAP}

\def\rhoc{\rho_{\rm c}}
\def\rhom{\rho_{\rm m}}

\def\mtom{M_{\rm 200m}}
\def\rtom{R_{\rm 200m}}

\def\mtoc{M_{\rm 200c}}
\def\rtoc{R_{\rm 200c}}

\newcommand{\mpch}{\>h^{-1}{\rm {Mpc}}}

\def\mvir{M_{\rm vir}}
\def\rvir{R_{\rm vir}}
\def\gcm3{\mathrm{g} / \mathrm{cm}^3}

\def\gtsima{$\; \buildrel > \over \sim \;$}
\def\ltsima{$\; \buildrel < \over \sim \;$}
\def\prosima{$\; \buildrel \propto \over \sim \;$}
\def\gsim{\lower.7ex\hbox{\gtsima}}
\def\lsim{\lower.7ex\hbox{\ltsima}}
\def\simgt{\lower.7ex\hbox{\gtsima}}
\def\simlt{\lower.7ex\hbox{\ltsima}}
\def\simpr{\lower.7ex\hbox{\prosima}}

\newcommand{\Mov}[1]{{\color{black}{#1}}}

\begin{document}

\title{Splashback radius in a spherical collapse model}

\author{Antonino~\surname{Del Popolo}}%
\affiliation{%
Dipartimento di Fisica e Astronomia, University of Catania, Viale Andrea Doria 6, 95125, Catania, Italy
}
\affiliation{%
Institute of Astronomy, Russian Academy of Sciences, 119017, Pyatnitskaya str., 48 , Moscow
}
\email[]{adelpopolo@oact.inaf.it}

\author{Morgan~\surname{Le Delliou}}%
\affiliation{Institute of Theoretical Physics, School of Physical Science and Technology, Lanzhou University, No.222, South Tianshui Road, Lanzhou, Gansu 730000, China}
\affiliation{Instituto de Astrof\'isica e Ci\^encias do Espa\c co, Universidade de Lisboa, Faculdade de Ci\^encias, Ed. C8, Campo Grande, 1769-016 Lisboa, Portugal}
\affiliation{Lanzhou Center for Theoretical Physics, Key Laboratory of Theoretical Physics of Gansu Province, Lanzhou University, Lanzhou, Gansu 730000, China}
\email[Corresponding author: ]{(delliou@lzu.edu.cn,) Morgan.LeDelliou.ift@gmail.com}

\label{firstpage}

\date{\today}

\begin{abstract}
It has been shown 
some years ago that dark matter haloes outskirts are characterized by very steep density profiles in a very small radial range. This feature has been interpreted as a pile up of at a similar location of different particle orbits, namely splashback material at half an orbit after collapse. Adhikari et al. (2014), obtained the location of the splashback radius through a very simple model, namely calculating a dark matter shell trajectory in the secondary infall model while it crosses a growing, NFW profile shaped, dark matter halo.
 Since they imposed a halo profile instead of calculating it from the trajectories of the shells of dark matter, they were not able to find the dark matter profile around the splashback radius. In the present paper, we use an improved spherical infall model taking into shell crossing, and several physical effects like ordered, and random angular momentum, dynamical friction, adiabatic contraction, etc. This allow us to determine the density profile from the inner to outer region, and study the behavior of the outer density profile. We will compare the density profiles, and the logarithmic slope of the density profile with the results of Diemer \& Kravtsov (2014) simulations, finding a good agreement between the prediction of the model and the simulations.

%
%
\end{abstract}

\pacs{98.52.Wz, 98.65.Cw}

\keywords{Dwarf galaxies; galaxy clusters; modified gravity; mass-temperature relation}

\maketitle

\section{Introduction}

The problem of determining the structure of dark matter haloes is an old one, and it has been studied from the analytical, numerical and observational point of view. The first efforts have been based on analytical models, and in particular on the spherical collapse model. The first trials to study the formation of virialized structure goes back to the seminal paper by 
\citet{Gunn1972}, after which several authors investigated the consequences of secondary infall, and accretion onto proto-structures, studying in particular the structure of the density 
profiles. \citet{Fillmore1984} studied the self-similar collapse of scale-free perturbations determining important processes in the halo profiles. Several other papers have improved on those previous results \citep[e.g.,][]{Hoffman1985,DelPopolo1997,Ascasibar2004,Williams2004,Hiotelis2006,DelPopolo2009,Cardone2011a}. 
These and other studies, improved the secondary infall model (SIM) taking into account the effects of ordered and random angular momentum, adiabatic contraction of dark matter (DM) produced by baryons \citep{Blumenthal1986,Gnedin2004,Gustafsson2006}, the effects of dynamical friction, and many more effects. Further development were obtained by means of DM only N-body simulations with the determination of the so called Navarro-Frenk-White profile \citep{Navarro1996,Navarro1997}, or of the Einasto profile \citep{Merrit2005,Navarro2010}, or by means of hydrodynamic simulations \citep{DiCintio2014a,Freundlich2020}.
Much of the past work has been focused on the inner structure of haloes, driven by the efforts to understand and solve the cusp-core problem, namely the discrepancy between the steep slopes predicted by DM only simulations, and the flat cored profiles observed in LSB (Low Surface Brightness), and dwarf galaxies. In the last years, some authors 
\citep[e.g.][]{Diemer2014} started to study the outer part of haloes, finding that the outer profiles are inconsistent with typical fits 
like the NFW 
or 
Einasto profiles. They found that the outer density profiles are characterized, over a narrow range of radii, by very steep logarithmic slopes, $d \log{\rho}/d \log{r} \leq -4$. 
According to \citet{Diemer2014}, the observed local steepening is due to a caustic, related to the splashback of material accreted by the halo. The presence of caustics is due to the pile up at a similar location of different particle orbits. The splashback radius correspond to the outermost caustic associated with the first apoapse after collapse. As shown in \citep{Adhikari2014}, the splashback location is given by the relation $d \log{\rho}/d \log{r} \leq -3$. 
Caustics are not a rare phenomenon. They are even present in the \citet{Fillmore1984} similarity solutions, in 3D similarity solutions of the collapse of triaxial peaks \citep{Lithwick2011}, or in real galaxies as radial shells \citep{Cooper2011}. However, to detect the density enhancement related to caustics is not easy in N-body simulations dark haloes. This is mainly due to the presence of small-scale structure smearing out caustics \citep{Diemand2008,Vogelsberger2011}.    
As material accumulates there, at the splashback radius is observed a steepening in the outer profile.  
A few years ago, \citet{Adhikari2014} estimated the location of the splashback radius by using a very simplified secondary collapse model, obtaining the secondary infall trajectory of DM shells by means of a growing DM halo with an NFW profile. In their calculation, a shape of the halo profile was imposed, instead of being computed from the trajectories of the DM shells. As a consequence in their calculation they were not able to obtain the full shape of the DM profile around the splashback radius. In the following, we introduce a much more complex spherical collapse model which will allow us to calculate simultaneously the trajectories and the DM halo profile. 

The paper is organized as follows. In Sec.~\ref{sec:model}, we discuss the model that will be used to determine the density profile, and will allow us the determination of the features of the outer profile. In Sec.~\ref{sec:numerical:definitions}, we discuss the results coming out from the model, and compare them with the results of \citet{Diemer2014}'s simulations. Sec.~\ref{sec:conclusions} is devoted to discussions.

\section{Model}\label{sec:model}

In this section, we discuss the model that will allow us the determination of the density profile. 
It was first introduced in \cite{DelPopolo2009}, followed by several applications: to density profiles universality studies \citep{DelPopolo2010,DelPopolo2011}, to galaxies \citep{DelPopolo2012a,DelPopolo2014} and clusters \citep{DelPopolo2012b,DelPopolo2014} density profiles, and to galaxies' inner surface-density distributions \citep*{DelPopolo2013d}.

The semi-analytical model (SAM) used in this work encompasses several upgrades on the SIM 
\citep[e.g.,][]{Gunn1972,Hoffman1985,DelPopolo1997,Ascasibar2004,Williams2004,Hiotelis2006,Cardone2011a,
DelPopolo2013a,DelPopolo2013b}. In contrast to anterior avatars of the SIM, it comprises non-radial collapse effects from 
random angular momentum \citep[RAM][]{Ryden1987,Gurevich1988a,Gurevich1988b,White1992,Sikivie1997,Nusser2001,Hiotelis2002,  
LeDelliou2003,Ascasibar2004,Williams2004,Zukin2010}
\footnote{RAM results from random velocities in the self-gravitating object \citep{Ryden1987}.}, ordered, tidal angular momentum \citep{Peebles1969,White1984}, the impact of dynamical friction \citep[e.g.,][]{AntonuccioDelogu1994,ElZant2001,ElZant2004,DelPopolo2009} and of baryonically induced DM adiabatic contraction
\citep{Blumenthal1986,Gnedin2004,Gustafsson2006}.

This SAM evolves perturbations from their linear expansion with the Hubble flow, reaching their turn-around, before collapsing with adiabatic central potential variations including shell-crossing" \citep{Gunn1977,Fillmore1984a}.

Spherical SIMs\footnote{An overdense perturbation sphere, within the homogeneous background Universe, provides a useful non-linear model (top hat toy model). Birkhoff's theorem is usually invoked to argue that such top hat overdensity collapses exactly as a closed sub-universe. Newtonian approach stands on stronger justification with Gau\ss~theorem. A more refined model divides the sphere into spherical ``shells", defined as the set of particles sharing the same orbit phase at a given radius \citep[see][]{LeDelliou2003}.} in the filiation from \citet{Gunn1972} describe a bound mass shell expansion from comoving initial radius $x_i$ to its maximum (turn-around or apapsis) radius $x_m=x_{ta}$,
\begin{equation}
x_m=g(x_i)=x_i/{\overline \delta_i},
\label{eq:pee}
\end{equation} 
with the linearly grown mean overdensity inside the shell extrapolated at current epoch $t_0$, with $\delta(y)$ obtained in Appendix B of \cite[Eq.~(B4)]{DelPopolo2009}, results from 
\begin{equation}
{\overline \delta_i}=\frac{3}{x_i^3} \int_0^{x_{i}} \delta(y)y^2 dy.
\label{eq:overd}
\end{equation}
More generally \citep{Peebles1980}, Eq.~\eqref{eq:pee} extend to a pure dust Universe, with density parameter $\Omega_i$, as
\begin{equation}
x_m=g(x_i)=x_i\frac{1+{\overline \delta_i}}{{\overline \delta_i}-(\Omega_i^{-1}-1)}.
\label{xtr}
\end{equation}
This generalisation represents the core of the SIM, for which a Lagrangian shell's time averaged radius remains proportional to its initial radius. Using Eq.~\eqref{xtr}, the final radius, $x$, can be written as proportional to the turn around radius, $x_m$
\begin{equation}
x=f(x_i) x_m
\label{eq:rc} 
\end{equation}
with the scaling fitted by \citep{Zaroubi1996}
\begin{align}
f = f(\alpha) =~& 0.186+0.156 \alpha+0.013 \alpha^2+0.017 \alpha^3 \nonumber\\&-0.0045 \alpha^4
+0.0032 \alpha^5.
\end{align}
Mass conservation at turn around radius yields the density profile \citep{Peebles1980,Hoffman1985,White1992}
\begin{equation}
\rho_{ta}(x_m)=\rho_i (x_i) \left( \frac{x_i}{x_m} \right)^2 \frac{d x_i}{dx_m}.
\label{eq:dturn}
\end{equation}
Beyond turnaround, shell crossing effects are bypassed with the Virial theorem, yielding a collapse factor $f=0.5$, resulting in the final density profile through mass conservation 
\begin{equation}
\rho(x)x^2 dx=\rho_i x_i^2 d x_i.
\label{eq:dturnn}
\end{equation}
This produces the power-law density profile from \citep{Hoffman1985} from the shell's initial density approximation
\begin{equation}
\rho_i(x_i)=\rho_{b,i} [1+ \delta_i(x_i)],
\end{equation}
and the linear $\delta_i$ expansion of Eq.~\eqref{xtr}, reading
\begin{equation}
\rho(x) \propto x^{-3(n+3)/(n+4)}.
\end{equation} 
However, the Virial theorem relies on energy conservation, and oscillations of collapsing shells through their inner shells break up the energy integral of motion and vary the value of $f$. 

This modifies the SIM dynamics, which assumed a ``gentle" collapse, as follows: with the conjecture of adiabatic variation of the central potential \citep{Gunn1977,Fillmore1984}, shells near the centre oscillate many times without significant changes in the potential. In other words, inner shells orbital period can be neglected compared with outer shells collapse time \citep{Zaroubi1993}. In this case, the inner shells admit the radial action $\oint v(r) dr$, with the radial velocity $v(r)$, as adiabatic invariant. The collapse of outer shells slowly changes the potential, shrinking the inner shells via the radial action invariant. 

The mass inside a shell with initial radius $x_i$, at its apocenter (apapsis radius) $x_m$ can be decomposed in the sum of its inner shells masses, i.e. with apocenters inside $x_m$, the permanent component $m_p$, with the contribution from its temporarily crossing outer shells masses temporarily, the additional mass $m_{add}$. Mass conservation yields the first component
\begin{equation}
m_p(x_m)=m(x_i)=\frac{4}{3} \pi \rho_{b,i} x_i^3 (1+{\overline \delta_i}),
\label{eq:mp}
\end{equation}
with initial time constant density $\rho_{b,i}$ of the homogeneous Universe. From Eq.~\eqref{eq:dturn}, the distribution of mass $m(x)=m(x_m)$ and the system radius $R$, its outer shell apapsis, together with the probability $P_{x_m}(x)$ to find the shell with apapsis $x$ inside the radius $x_m$, follows the additional component
\begin{equation}
m_{add}(x_m)=\int_{x_m}^{R} P_{r_m} (x) \frac{d m(x)}{dx} dx,
\label{eq:madd}
\end{equation}
the resulting total mass reading
\begin{equation}
m_T(x_m)=m_p(x_m)+m_{add}(x_m).
\label{eq:mpp}
\end{equation}
The ratio of the time spent below $x_m$ to the outer shell $x$ oscillation period allows to compute $P_{x_m}(x)$ from the outer shell's pericenter $x_p$ and $v_x (\eta)$, its radial velocity at radius $\eta$, as
\begin{equation}
P_{x_m}(x)= \frac{
\int_{x_p}^{x_m} \frac{d \eta}{v_x(\eta)} }
{\int_{x_p}^{x} \frac{d \eta}{v_x(\eta)}
}.
\end{equation}
The outer shell's radial velocity derives from integrating its equation of motion, including the tidal torques generated ordered specific angular momentum\footnote{Defined in \citep[Appendix B]{DelPopolo2009}, with $\nu=\delta(0)/\sigma$, and $\sigma$, the mass variance averaged on a scale $R_f$.} $h(r,\nu )$, the random angular momentum $j(r, \nu)$\citep[see][and follow ups]{Ryden1987,DelPopolo2009},the gravitational potential acceleration $G(r)$, $\Lambda$ the cosmological constant and the dynamical friction coefficient $\mu$: 
\begin{equation}
\frac{dv_r}{dt}=\frac{h^2(r,\nu )+j^2(r, \nu)}{r^3}-G(r) -\mu \frac{dr}{dt}+ \frac{\Lambda}{3}r.
\label{eq:coll}
\end{equation}
Computations of $\mu$ and the angular momenta are explained in \citep[Appendices C and D]{DelPopolo2009}. In the restriction where $\mu=0$, Eq.~\eqref{eq:coll} integrates into the square of velocity:
\begin{equation}
v(r)^2=2 \left[\epsilon -G \int_0^r \frac{m_T(y)}{y^2} d y +\int_0^r \frac{h^2}{y^3} dy + \frac{\Lambda}{6} r^2 
\right],
\label{eq:FrictionlessV}
\end{equation}
where the shell's specific binding energy results from the turnaround value at $v(r)=dr/dt=0$ at $r=x_m$ in Eq.~\eqref{eq:FrictionlessV}.

The $\mu\neq 0$ case requires the numerical integration of Eq.~\eqref{eq:coll} for $v$, after its quadrature into 
\begin{equation}
\frac{d v^2}{d t}+2 \mu v^2=2 \left[
\frac{h^2+j^2}{r^3} -G\frac{m_T}{r^2} + \frac{\Lambda}{3} r
\right] v.
\label{eq:veloc}
\end{equation}

With the computations above complete, following \citep{Gunn1977,Fillmore1984}, we obtain $f(x_i)$ the shell's collapse factor which starts at radius $x_i$ and reaches apapsis $x_m$
\begin{equation}
f(x_i)=\frac{m_p(r_m)}{m_p(r_m)+m_{add}(r_m)},
\label{eq:cfact}
\end{equation}
and, from Eqs.~\eqref{eq:dturn} and \eqref{eq:rc}, the corresponding density profile at Virialisation
\begin{equation}
\rho(x)=\frac{\rho_{ta}(x_m)}{f^3} \left[1+\frac{d \ln f}{d \ln g} \right]^{-1}.
\label{eq:dturnnn}
\end{equation}

The calculations above allow to evaluate the $f(x_i)$ variations from energy integral break up, confirmed from N-body simulations, finding its relation to the initial density perturbation profile and its increase with initial radius. The case $x_i\rightarrow 0$ and $f \rightarrow 0$ recovers a radial collapse \citep{Lokas2000}. The computation from Eq.~\eqref{eq:cfact} of $f$ via integration of Eq.~\eqref{eq:madd} can proceed numerically after variable change to express them in terms of initial radius \citep{Lokas2000,Hiotelis2002}. Such variable change turns Eq.~\eqref{eq:madd} into
\begin{equation}
m_{add}(r_m)=4\pi\rho_{b,i}\int_{x_i}^{x_b}P_{x_i}(x'_i)[1+\delta_i(x'_i)]x_i'^2\mathrm{d}x'_i,
\label{eqb10}
\end{equation}
with $P_{x_i}(x'_i)=I(x_i)/I(x'_i)$,
\begin{equation}
I(r)=\int_{x'_p}^r\frac{1}{v_{g(x'_i)}(g(\eta))}\frac{\mathrm{d}g(\eta)}{\mathrm{d} \eta}\mathrm{d} \eta,\\
\label{eqb11}
\end{equation}
taking $r_m = g(x_i)$, $x'_p=g^{-1}(x_p)$, and the initial $x'_i$ shell's pericenter $x_p$. The upper bound $x_b$ of Eq.~\eqref{eqb10}'s integration corresponds to the presently collapsed sphere's initial radius. A similar variable change in Eq.~\eqref{eq:veloc} leads to the determination, for a shell characterised with apapsis $x=g(x_i)$, of the radial velocity $v$ at radius $r=g(r_i)$ through
\begin{equation}
\frac{d v^2_x(r)}{d t}+2 \mu v_x^2=2 \left[
\frac{h^2_x+j^2_x}{r^3} - \Psi(r)
+ \frac{\Lambda}{3} r
\right] v_x(r),
\label{eq:vell}
\end{equation}
with the shifted gravitational potential $\Psi$, from the initial mass profile  $m(x_i)$, reading
\begin{equation}
\Psi[g(r_i)]=\frac{Gm(x_b)}{g(x_b)}+G\int_{r_i}^{x_b}\frac{m(x_i)}{g^2(x_i)}
\frac{\mathrm{d}g(x_i)}{\mathrm{d}x_i}\mathrm{d}x_i.
\end{equation}

In summary, \citep[see also][Sec.~4]{Lokas2000} the equation of motions of a shell \eqref{eq:vell}, given angular momentum distribution, dynamical friction coefficient and initial conditions, integrates to compute $P_{x_i}$, the probability from Eq.~\eqref{eqb11}, that then yields the transient part of gravitating mass acting on the shell $m_{add}$, from Eq.~\eqref{eqb10}, the collapse factor $f$, from Eq.~\eqref{eq:cfact}, to obtain in the end the density profile through Eq.~\eqref{eq:dturnnn} \citep[also in][Sec. 2.1]{Ascasibar2004}. Our model's ordered angular momentum formation follows \citep[Appendix~C1]{DelPopolo2009}, its random angular momentum computation agrees with \citep[Appendix~C2]{DelPopolo2009}, its dynamical friction coefficient, and the baryon dissipative collapse, are described in \citep[Appendix~D]{DelPopolo2009}, baryon's adiabatic contraction is the object of \citep[Appendix~E]{DelPopolo2009}, while \citep[Appendix~B]{DelPopolo2009} describes initial condition generation. More explicitly, the model's "ordered angular momentum" $h$ derives from the tidal torque theory (TTT) 
\citep{Hoyle1953,Peebles1969,White1984,Ryden1988,Eisenstein1995}, which processes from the tidal torques exerted on smaller scales structures by larger scales objects. On another hand, the "random angular momentum" $j$ is computed from the orbital axis ratio, between the pericentric and apocentric radii $r_{\rm min}$ and  $r_{\rm max}$, noted $e=\left(\frac{r_{\rm min}}{r_{\rm max}}\right)$ \citep{AvilaReese1998}, modified according to the system's dynamical state, following simulations \cite{Ascasibar2004}, into
\begin{equation}
e(r_{\rm max})\simeq 0.8\left(\frac{r_{\rm max}}{r_{\rm ta}}\right)^{0.1}\;,
\end{equation}
a function of the ratio of $r_{\rm max}$ to the turnaround radius, with $r_{\rm max}<0.1 r_{\rm ta}$. As for the dynamical friction effects, they were introduced in the equation of motion with a dynamical friction force, as computed in \citep[][Appendix~A, see Eq.~A14]{DelPopolo2009}. Finally, for the density profile steepening from adiabatic compression, the methods of \cite{Gnedin2004} were followed.

\Mov{ 
\section{Population of haloes}

The model is the principle that transforms from initial conditions to charateristics of the galaxy. Generating a population of galaxies requires to use the model but also some initial range of parameters, a population of initial parameters in other words, that gives the statistical distribution of galaxies, given by a sufficiently large number of the generated population. Concerning the initial conditions and the determination of the density profile of galaxies 
 it is necessary to calculate the initial overdensity $\overline \delta_i (x_i)$. 
This can be calculated when the spectrum of perturbations is known. 
It is widely accepted that structure formation 
in the universe is generated through the growth and collapse 
of primeval density 
perturbations originated from quantum fluctuations in an inflationary 
phase of early Universe. The growth in time of small 
perturbations is due to gravitational instability. The statistics of 
density fluctuations originated in the inflationary era are Gaussian, and 
can be expressed entirely 
in terms of the power spectrum of the density fluctuations: 
\begin{equation}
P( k) = \langle |\delta_{{\bf k}}|^{2} \rangle 
\end{equation}
where 
\begin{equation}
\delta_{{\bf k}} =\int d^{3} k \,\exp(-i {\bf k x}) \delta({\bf x})
\end{equation}
\begin{equation}
\delta({\bf x}) = \frac{ \rho ({\bf x}) - \rho_{b}}{ \rho_{b} }
\end{equation}
and $ \rho_{b} $ is the mean background density. 
In biased structure formation theory it is assumed that cosmic structures 
of linear scale $ R_f$ form around the peaks of the density field, 
$  \delta( {\bf x})$, smoothed on the same scale. 
According to the hierarchical scenario of structure formation, 
haloes should collapse around maxima of the smoothed density field (see below).  
The statistics of peaks in a Gaussian random field has been studied in the classical paper by \citet[hereafter BBKS]{Bardeen1986}. 
A well known result is the expression for the radial density profile of a fluctuation centered on a primordial 
peak of arbitrary height $\nu$:
\begin{equation}
\langle \delta (r) \rangle =\frac{\nu \xi (r)}{\xi (0)^{1/2}}-\frac{\vartheta (\nu
\gamma ,\gamma )}{\gamma (1-\gamma ^2)}\left[ \gamma ^2\xi (r)+\frac{%
R_{\ast }^2}3\nabla ^2\xi(r) \right] \cdot \xi (0)^{-1/2} 
\label{eq:dens}
\end{equation}
\cite[with \cite{Ryden1987} hereafter RG87]{Bardeen1986,Ryden1987},
where $\nu= \delta(0)/\sigma $ (see the following for a definition of $\sigma$) is the height of a density peak, $\xi (r)$ is the two-point 
correlation function:
\begin{equation}
\xi(r)= \frac{1}{2 \pi^2 r} \int_0^{\infty} P(k) k \sin(k r) d k
\end{equation}
$\gamma $ and $R_{\ast}$ are two spectral parameters
given respectively by:
\begin{equation}
\gamma =\frac{\int k^4P(k)dk}{\left[ \int k^2P(k)dk\int k^6P(k)dk\right]
^{1/2}}
\label{eq:gammm}
\end{equation}
\begin{equation}
R_{*}=\left[ \frac{3\int k^4P(k)dk}{\int k^6P(k)dk}\right] ^{1/2}
\label{eq:rrr}
\end{equation}
while $ \vartheta (\gamma \nu ,\gamma )$ is: 
\begin{equation}
\vartheta (\nu \gamma ,\gamma )=\frac{3(1-\gamma ^2)+\left( 1.216-0.9\gamma
^4\right) \exp \left[ -\left( \frac \gamma 2\right) \left( \frac{\nu \gamma }%
2\right) ^2\right] }{\left[ 3\left( 1-\gamma ^2\right) +0.45+\left( \frac{%
\nu \gamma }2\right) ^2\right] ^{1/2}+\frac{\nu \gamma }2}
\label{eq:tet}
\end{equation}

Then $\overline \delta_i$ is calculated from Eq. (\ref{eq:dens}) similarly to \citet[their Section 2.2]{Ascasibar2004}.
In order to calculate $\delta(r)$ we need a power spectrum, $P(k)$.
The CDM spectrum used in this paper is that of BBKS, with transfer function:
\begin{align}
T(k) =& \frac{[\ln \left( 1+2.34 q\right)]}{2.34 q}
\cdot [1+3.89q+
(16.1 q)^2+(5.46 q)^3\nonumber\\
&+(6.71)^4]^{-1/4}
%
%
\label{eq:ma5}
\end{align}
where 
$q=\frac{k\theta^{1/2}}{\Omega_{\rm X} h^2 {\rm Mpc^{-1}}}$.
Here $\theta=\rho_{\rm er}/(1.68 \rho_{\rm \gamma})$
represents the ratio of the energy density in relativistic particles to
that in photons ($\theta=1$ corresponds to photons and three flavors of
relativistic neutrinos). The spectrum is connected to the transfer function through the equation:
\begin{equation}
P(k)=P_{CDM} e^{-1/2 k^2 R_f^2}
\end{equation}
where $R_f$ is the smoothing (filtering) scale and $P_{CDM}$ is given by:
\begin{equation}
P_{CDM}= A k T^2(k)
\end{equation}
where $A$ is the normalization constant. 
We normalized the spectrum by imposing that the mass variance of the density field
\begin{equation}
\sigma^2(M)=\frac{1}{2 \pi^2} \int_0^\infty dk k^2 P(k) W^2(kR)
\end{equation}
convolved with the top hat window 
\begin{equation}
W(kR)=\frac{3}{(kR)^3} (\sin kR-kR \cos kR)
\end{equation}
of radius 8 $h^{-1}$ $Mpc^{-1}$ is $\sigma _{8}=0.76$. 
%
Throughout the paper we adopt a $\Lambda$CDM cosmology with WMAP3 parameters \cite{Spergel2007}, $\Omega_m=1-\Omega_{\Lambda}=0.24$,  $\Omega_{\Lambda}=0.76$, $\Omega_b=0.043$ and $h=0.73$, where $h$ is the Hubble constant in units of 100 km $s^{-1}$ $Mpc^{-1}$. 



The mass enclosed in $R_f$ is calculated, as in RG87, as $M=4 \pi/3 \rho_b R_f^3$, so that for $R_f=0.12$ Mpc, $M \simeq 10^9 M_{\odot}$
\footnote{For precision sake, the mass scale $M$ is connected to the smoothing scale by:
$M_G=(2 \pi) ^(3/2) \b R_{f}^3$
for a Gaussian smoothing ($P(k,R_f)=e^{-R_f^2 k^2}P(k)$) and by
$M_{TH}=4 \pi/3 \b R_{TH}^3$
for top hat smoothing.
The mass enclosed by the smoothing function applied to the uniform background is the same for $R_f=0.64 R_{TH}$ (see BBKS). 
}. 
Structure like Galaxies form from high peaks in the density field, high enough so that they stand out above the 
``noise" and dominate the infall dynamics of the surrounding matter. 

The amplitude of any given peak is expressed in terms of its $\sigma$ deviation, where 
$\sigma=\xi(0)^{1/2}$.  
Thus the central density contrast of an $ \nu \sigma$ peak is $ \nu \xi(0)^{1/2}$ and the peak height is given by $\nu=\delta(0)/\sigma$.
Given that galaxies are rather common, they must have formed from 
peaks that are not very rare, say, 2-4 $\sigma$ peaks (RG87). 
In \citet[][Fig. 6 of Appendix B]{DelPopolo2009} is plotted the density profiles $\delta(x)$ for 
$\nu=2,3,$ and $4$. 
We generated a set of galaxies starting from the initial conditions and using the model. The different masses are related to the filtering radius $R_f$, and the peak height $\nu$. 
As seen in \cite[Fig. 6 of Appendix B,][]{DelPopolo2009}, the value of $\delta(x)$, and then the final density profile changes with changing $\nu$. The halo characteristics are also modified by the tidal torque (ordered angular momentum), the random angular momentum \cite[Eq.~C17 in][]{DelPopolo2009}, dynamical friction \cite[Appendix D in][]{DelPopolo2009}, and adiabatic compression \cite[Appendix E in][]{DelPopolo2009}. 
}

\Mov{In summary, the statistics of our model's halo populations is given by the BBKS Gaussian random field fluctuations. The sample size generated is the result of computing power constraints versus stability of the median density profile, and was set after verifying that an increase in sample size would not significantly affect the results.
}

\section{
Results
}
\label{sec:numerical:definitions}

Before describing the results, we give here some definitions that are commonly been 
used. The three dimensional radius with respect to the center of the halo is indicated with $r$, while the capital $R$ is used to define radii related to the mass of the halo. The critical density is indicated by $\rho_{\rm c}$, and the mean matter density with 
$\rho_{\rm m}$. Masses at given overdensity are given by $M_{\Delta \rm m} = M(<R_{\Delta \rm m})$. For example the mass with overdensity at $\Delta=200$ is $\mtom$, while that corresponding to the critical density $\mtoc$ reads $M_{\Delta \rm c} = M(<R_{\Delta \rm c})$. $\mvir$ and $\rvir$ are related to $\Delta(z)$, which, at $z=0$ corresponds to $\Delta_{\rm vir}(z=0) \approx 358$. 
Instead of using masses, haloes are binned by mean values of the peak height $\nu$, since, at fixed $\nu$, halo properties across redshifts should likely be similar. 
The definition of the peak height follows the usual
\begin{equation}
\nu \equiv \frac{\delta_{\rm c}}{\sigma(M, z)} = \frac{\delta_{\rm c}}{\sigma(M, z = 0) \times D_+(z)},
\end{equation}
where the critical density is given according to the spherical top hat model by $\delta_{\rm c} = 1.686$. $D_+(z)$ represents the linear growth factor normalized to unity at $z=0$. Given a sphere of radius $R$, the rms density fluctuation is given by 
\begin{equation}
\sigma^2(R) = \frac{1}{2 \pi^2} \int_0^{\infty} k^2 P(k) |\tilde{W}(kR)|^2 dk.
\end{equation}
Here $W$ is the spherical top hat filter function, and $\tilde{W}(kR)$, its Fourier transform. 
The linear power spectrum $P(k)$ is given by means of the formula of 
\citet{Eisenstein1998}, with the normalization $\sigma(8 \mpch) = \sigma_8$ = $0.82$.
$\sigma(M) = \sigma(R[M])$ indicates the variance at a given mass, and the calculation of $\nu$ uses $M = \mvir$. The relation between $\nu$ and the virial mass is represented in \citep[Fig.~1]{Diemer2014}. Halo mass can be translated in $\nu$ by means of $\rvir$, 
because that radius corresponds to the largest for which, at fixed mass, the density profile scatter remains relatively small. This is also why we preferred $\rvir$ to $\rtom$ for computing the difference of mass accretion rate between two redshifts.
The mean or median profiles in rescaled radial units, are obtained rescaling the individual haloes profile using $R_{\Delta}$ of the halo. The mean and median are obtained from the rescaled profiles. The slope profiles are obtained using the fourth-order Savitzky--Golay smoothing algorithm over the 15 nearest bins \citep{Savitzky1964}, and the functional fits are obtained by means of the Levenberg-Marquart algorithm. Median profiles are used since they are a good approximation of the typical profile, and can be used to study trends in the density profiles. 

\begin{figure*}[!ht]
 \centering
\includegraphics[width=7cm,angle=0]{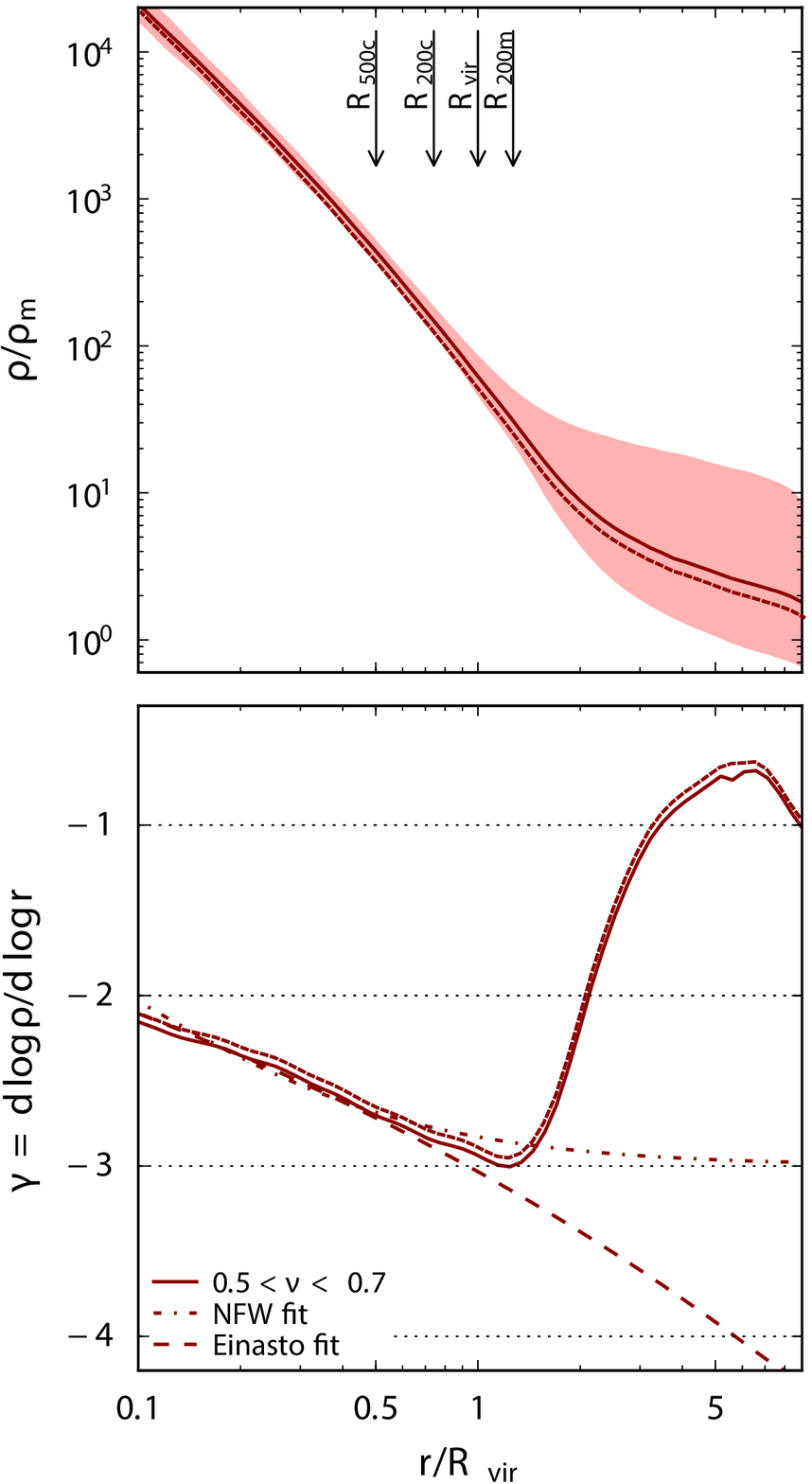}
\includegraphics[width=7cm,angle=0]{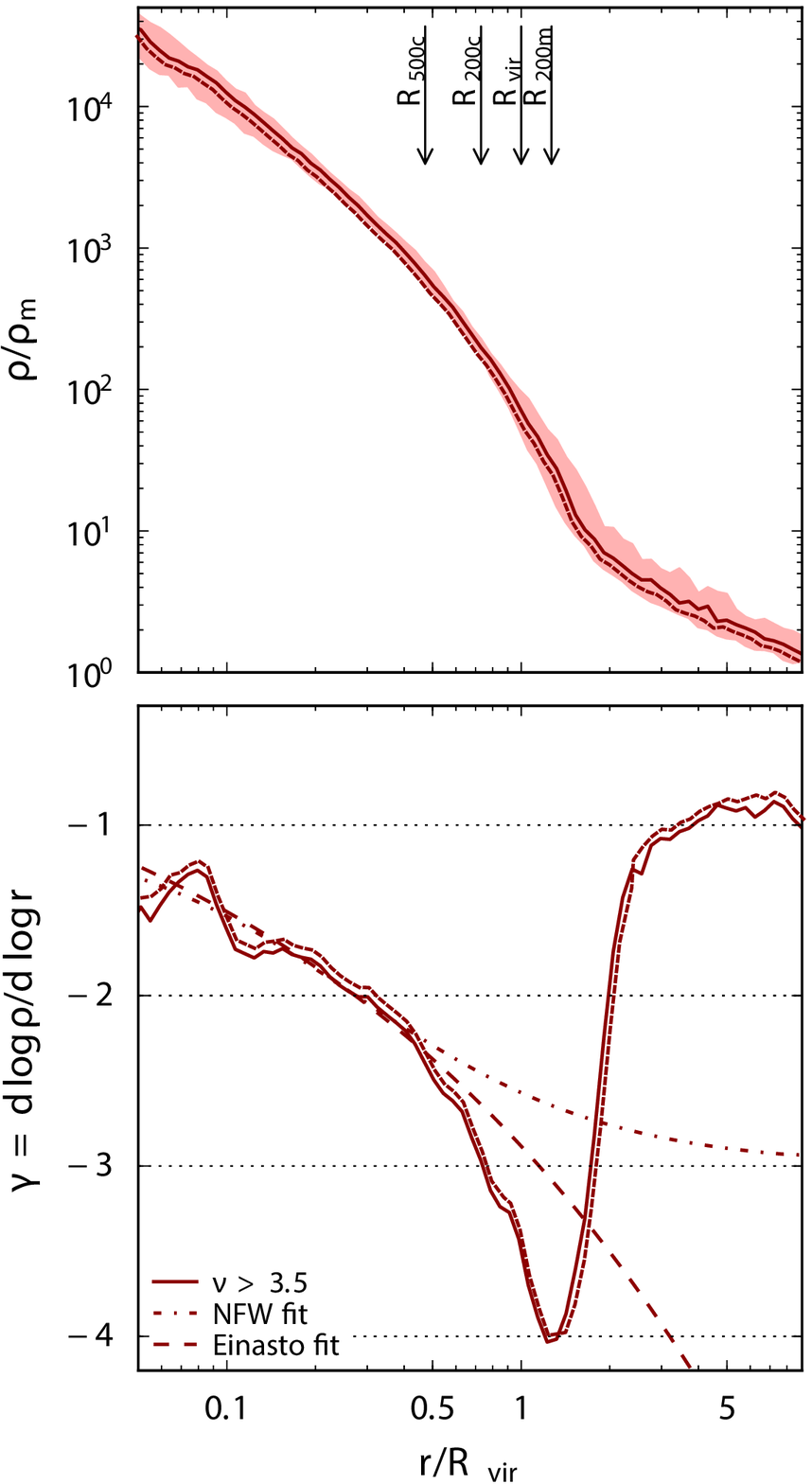}
 \caption[justified]{Median density profiles of low-mass (top left), and very massive haloes (top right) at $z=0$. The shaded band represents the interval around the median containing 68\% of the profiles of individual halos in the $\nu$ bin. In the bottom panel are represented the logarithmic slope profiles from the top panels profiles. Are also indicated the NFW and Einasto fits to the profiles. The solid lines represent the results from \cite{Diemer2014}. The dotted lines, in the top panels, represent the median density profiles, while the bottom panels show the logarithmic slope profile, both obtained with our model.}
 \label{fig:comparison1}
\end{figure*}

\begin{figure*}[!ht]
 \centering
\includegraphics[width=5cm,angle=0]{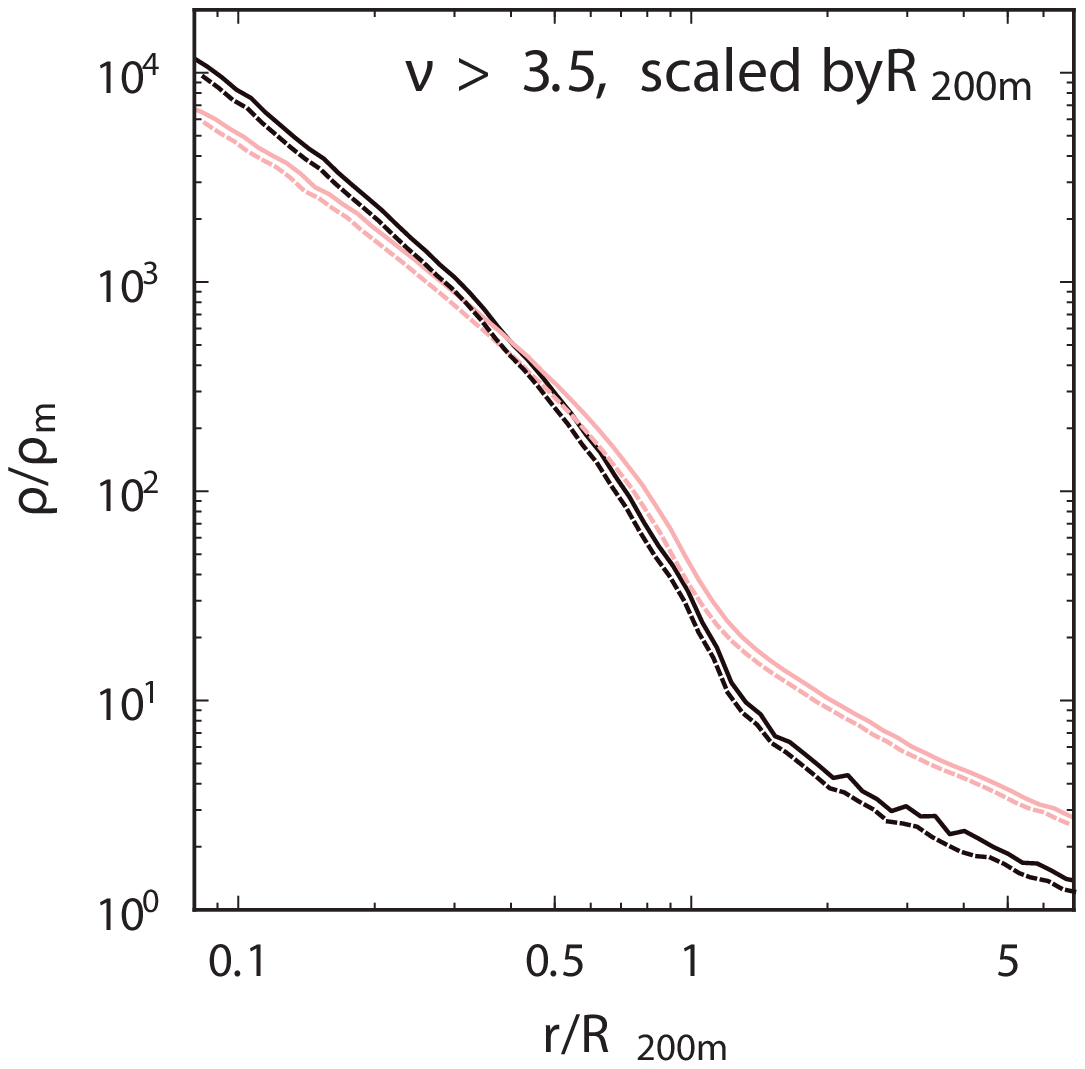}
\includegraphics[width=5cm,angle=0]{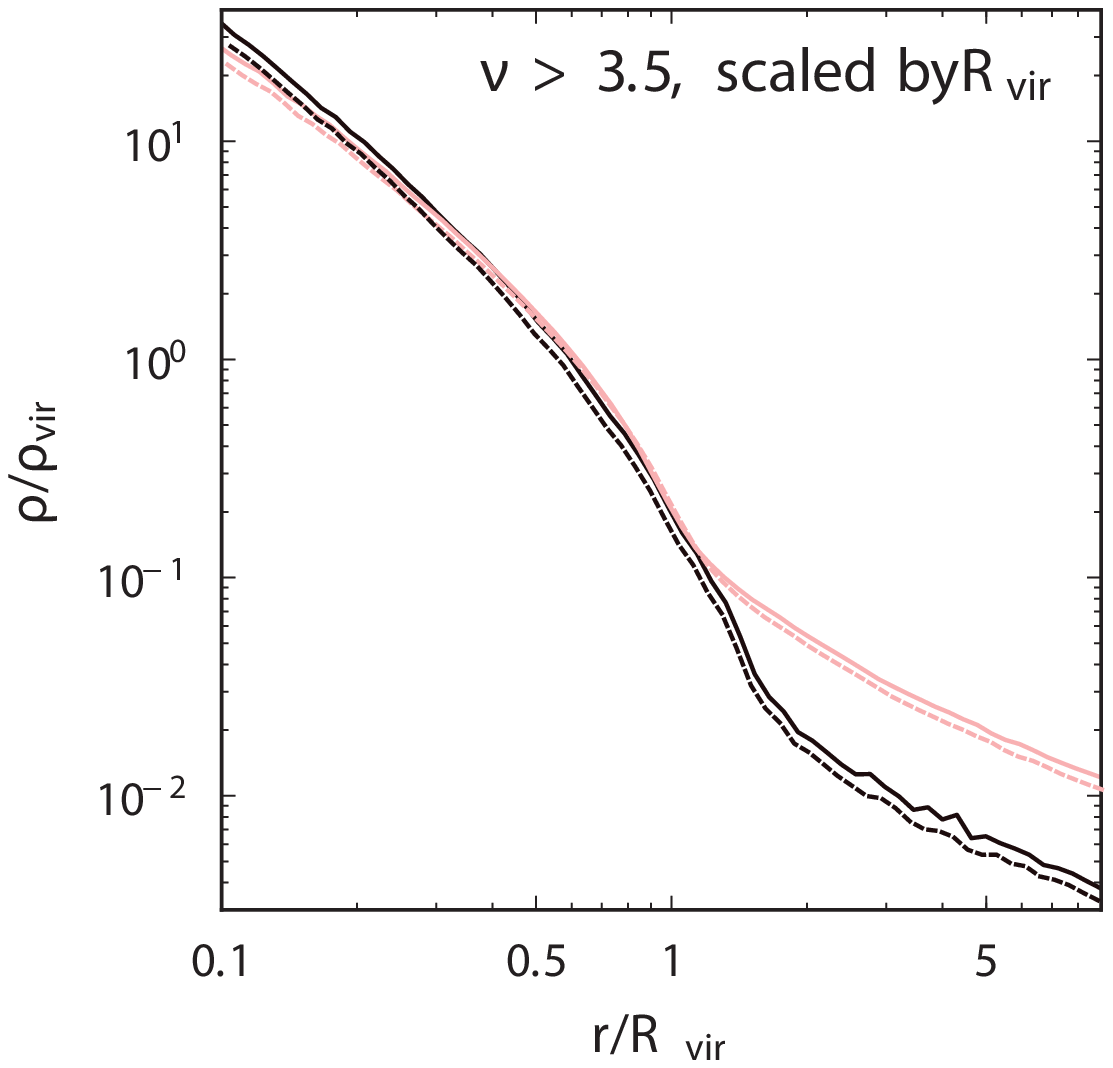}
\includegraphics[width=5cm,angle=0]{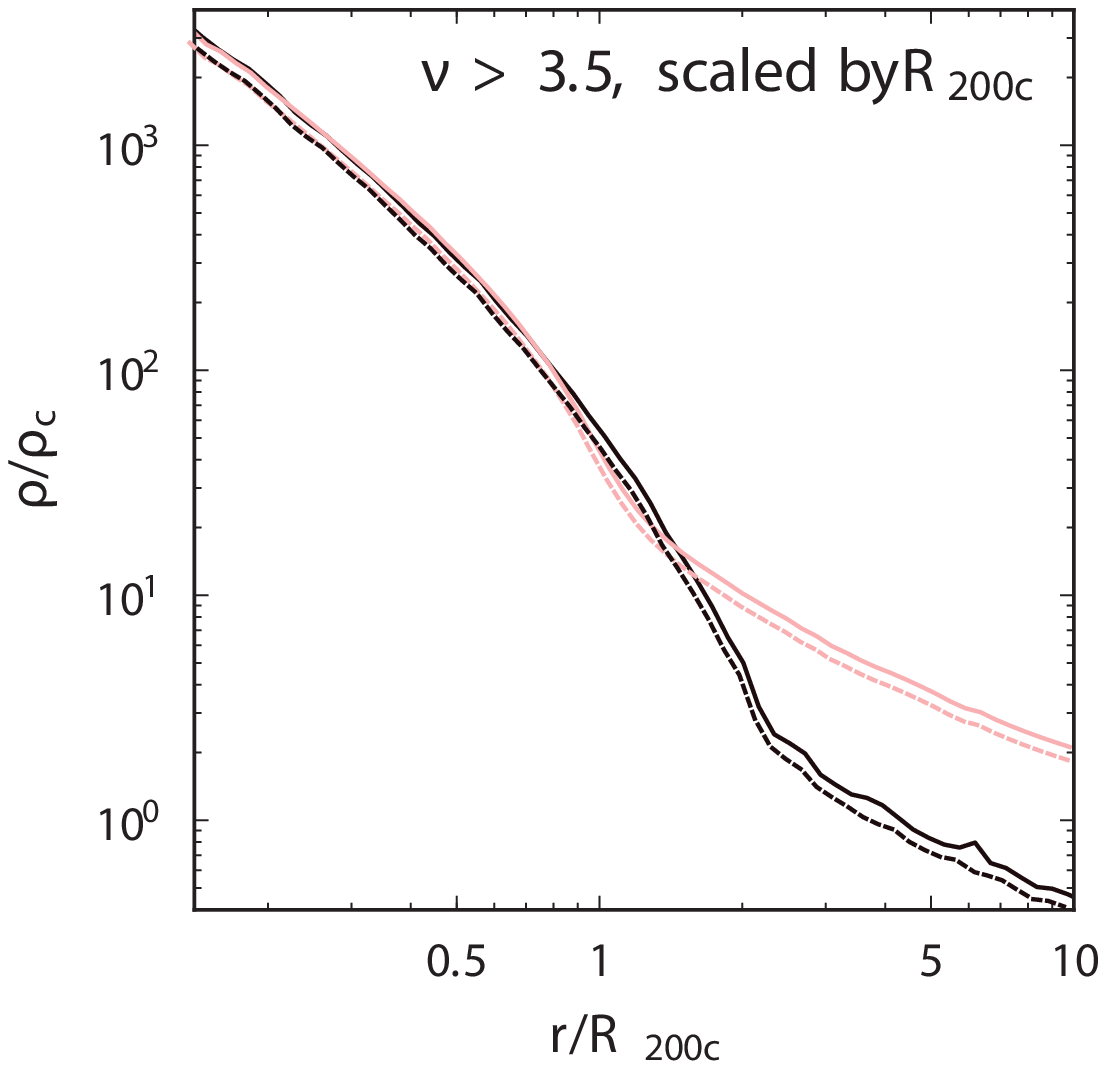}
 \caption[justified]{Self-similarity of the redshift evolution of the profiles. In all panels, solid lines represent the results from \cite{Diemer2014} while dotted lines show results obtained with our model.Left panel: redshift evolution of the median density profiles for a peak with $\nu >3.5$, as a function of the radius rescaled by $R_{200m}$, with density rescaled by $\rho_m$. Central panel: represents the same density profiles rescaled by $R_{\rm vir}$, and $\rho_{\rm vir}$, respectively. Right panel: same density profiles rescaled by $R_{\rm c}$, and $\rho_{\rm c}$, respectively. The black lines corresponds to $z=0$, and the red lines to $z=6$.}
 \label{fig:comparison2}
\end{figure*}

\begin{figure*}[!ht]
 \centering
\includegraphics[width=5cm,angle=0]{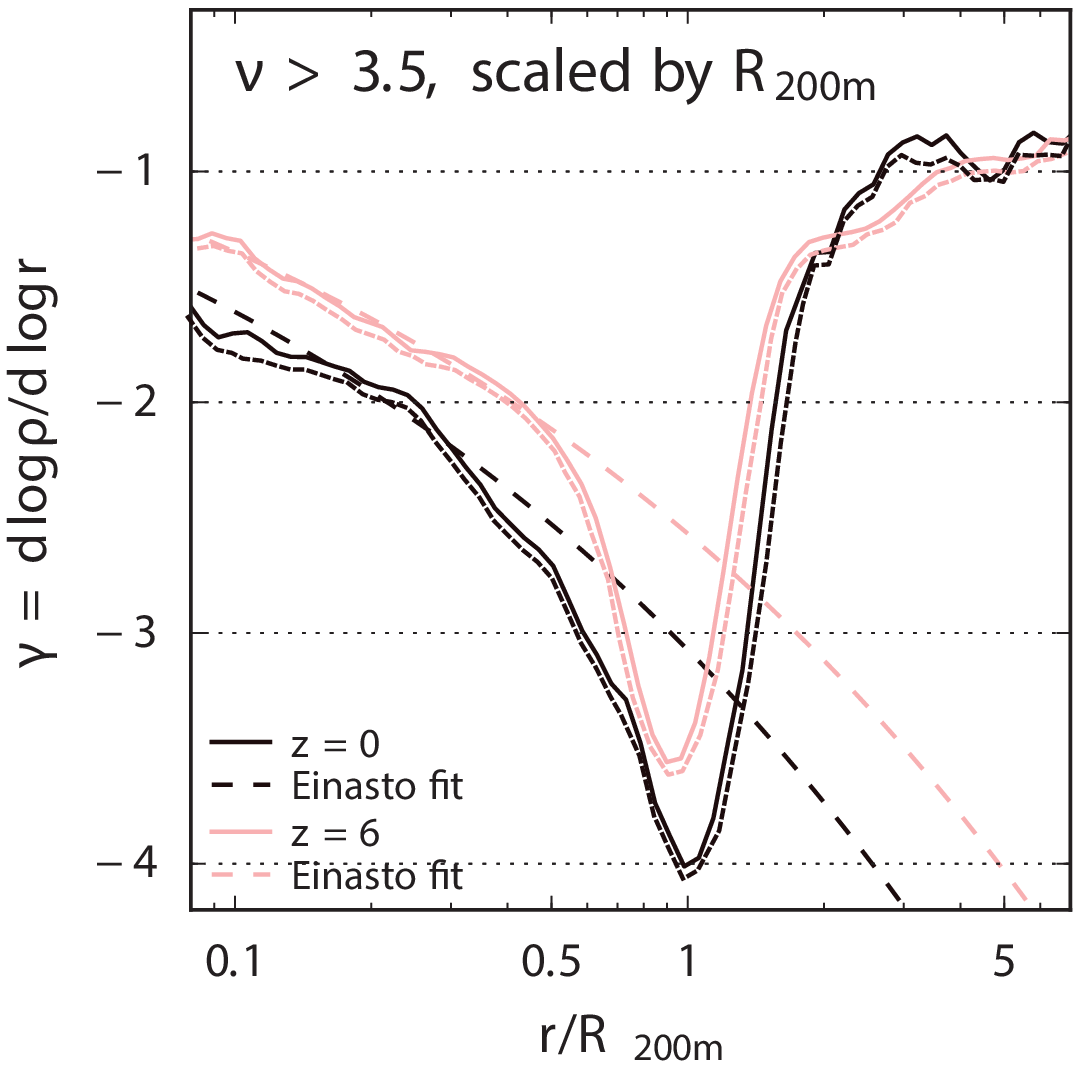}
\includegraphics[width=5cm,angle=0]{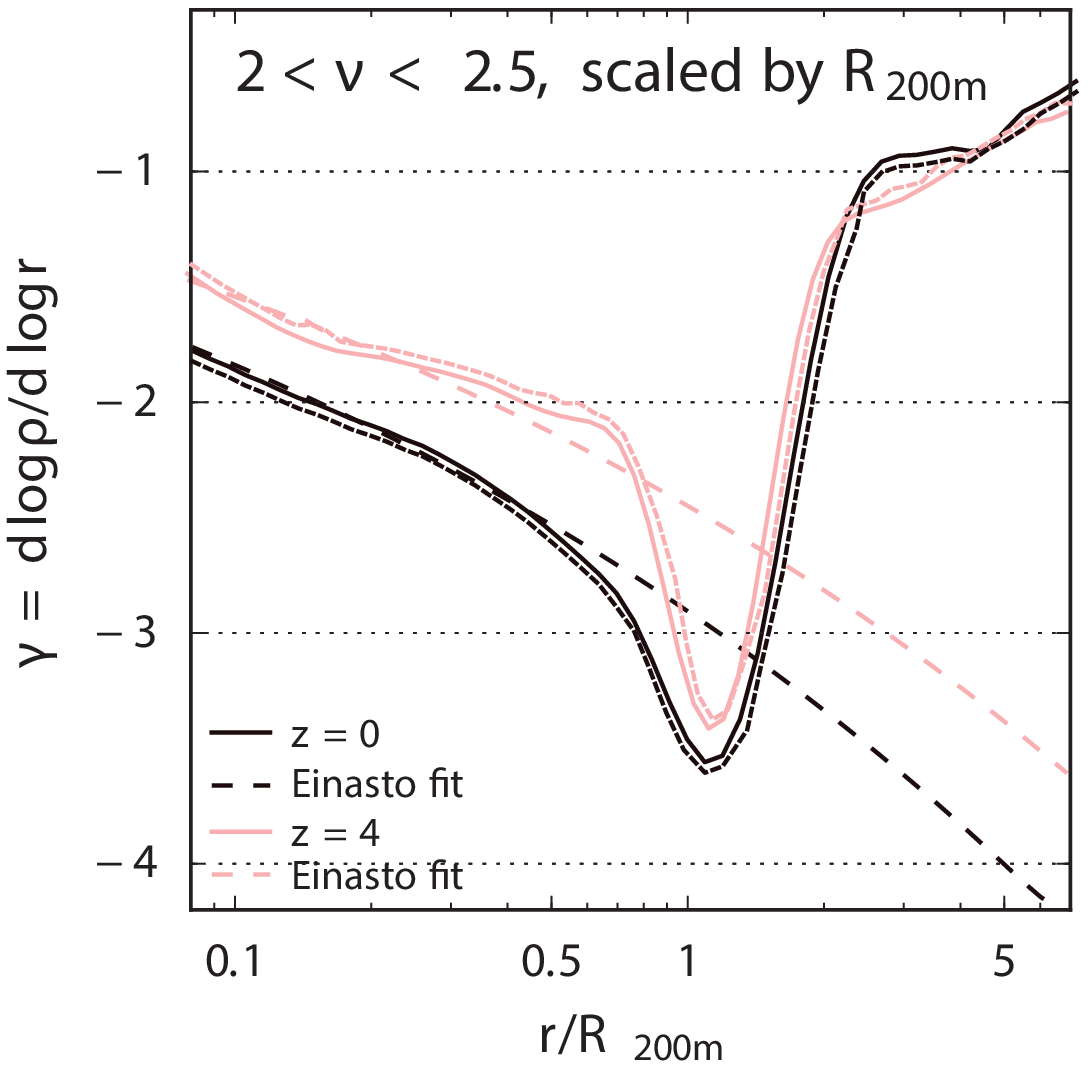}
\includegraphics[width=5cm,angle=0]{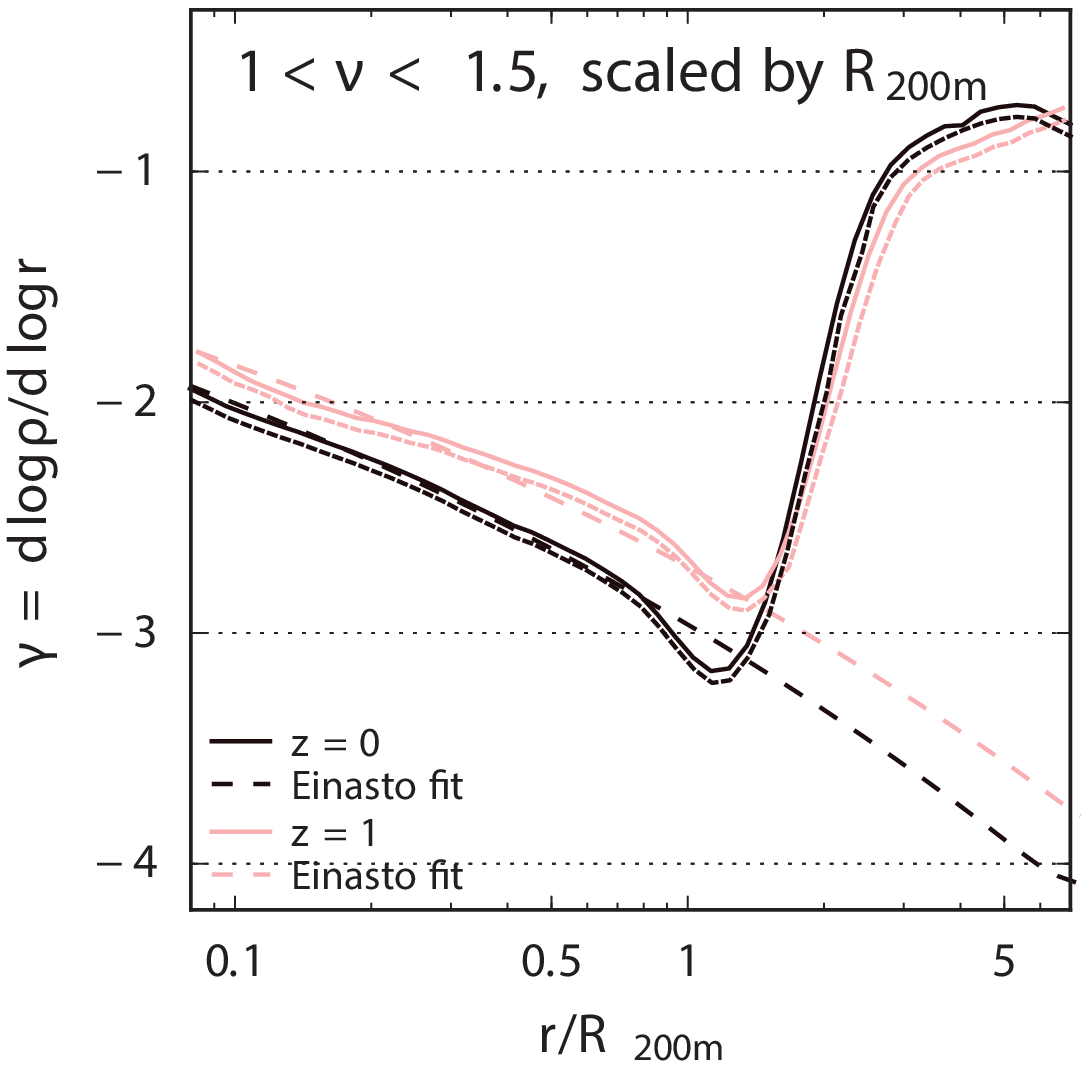}
\includegraphics[width=5cm,angle=0]{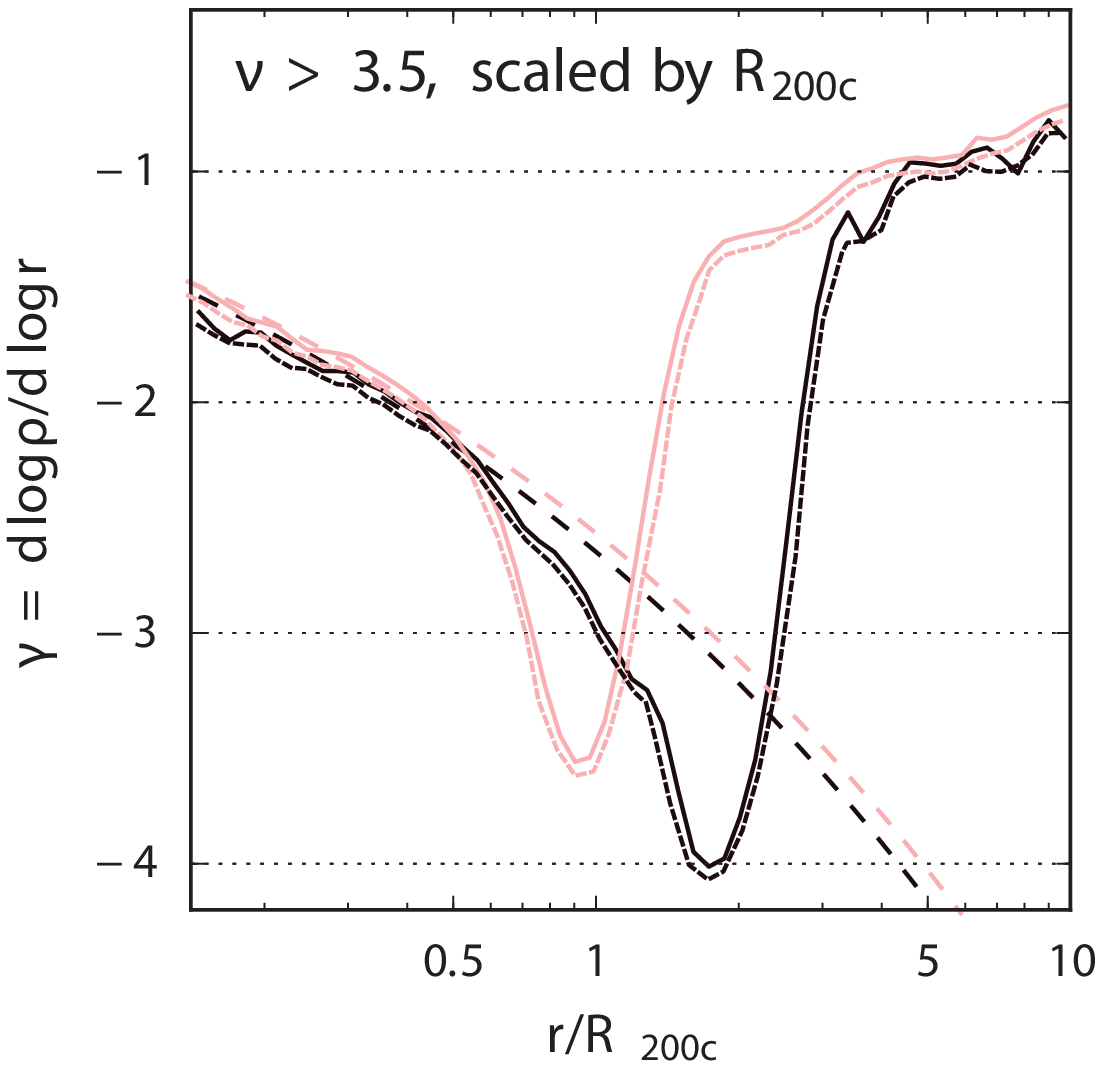}
\includegraphics[width=5cm,angle=0]{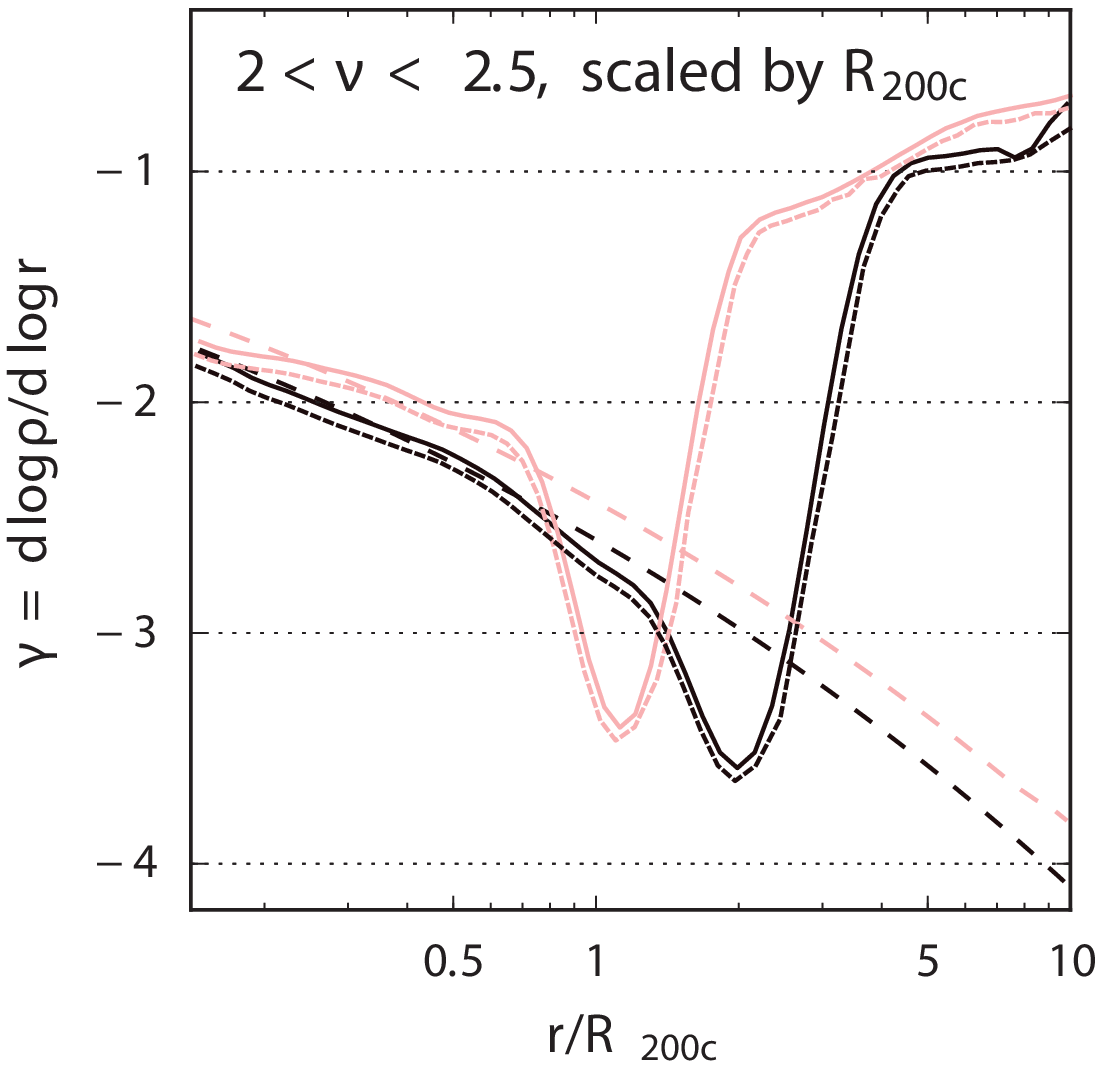}
\includegraphics[width=5cm,angle=0]{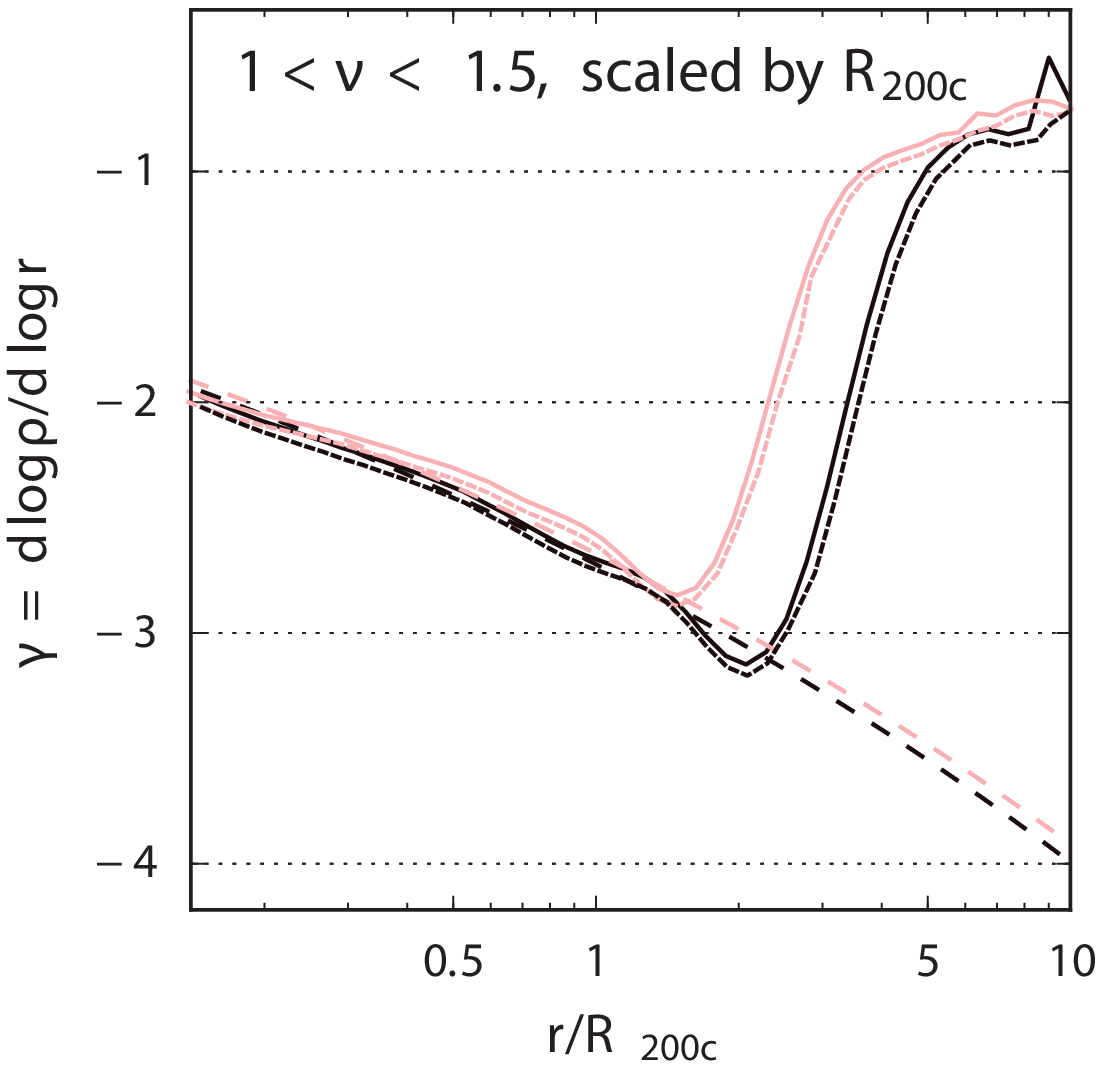}
\caption[justified]{Logarithmic slope of three $\nu$ bins at different redshifts. The left panels are related to the $\nu >3.5$ sample of Fig.~\ref{fig:comparison2}. The central, and right panels refer to the $\nu$ bins shown in \citep[Fig.~5]{Diemer2014}. Radii in the top panels are rescaled in units of $R_{\rm 200m}$, while those 
  in the bottom panel are expressed in units of $R_{\rm 200c}$. Again, the solid lines stand for the \cite{Diemer2014} simulations. The dashed lines provide fits using the Einasto profile. The dotted lines represents the result of our model. Black, and red lines corresponds to different redshifts indicated in the legends of the top panels.}
 \label{fig:comparison3}
\end{figure*}

In this paper, we use the model described in Sec.~\ref{sec:model} to generate a population of haloes from which to build up median density profiles of haloes that are binned by peak height.

We will compare the results of our model with those of the simulations of \citep{Diemer2014}. The goal is to show that the model is giving a correct description of the outer density profile of haloes.

In Fig.~\ref{fig:comparison1}, we plot the median density profiles of two different halo samples, from \cite{Diemer2014} (solid lines) or from our model (dotted lines), and their logarithmic slope profiles $\gamma(r) \equiv d \log \rho / d \log r$. 
The top panels show the median density profiles of low-mass (top left), and very massive haloes (top right) at $z=0$. The shaded band represents the interval around the median containing 68\% of the profiles of individual halos in the $\nu$ bin  from\cite{Diemer2014}. In the bottom panel are represented the logarithmic slope profiles corresponding to the top panels. Are also indicated the NFW and Einasto fits to the profiles. 
The low-mass sample (left panels) is characterized by $0.5 < \nu < 0.7$, while the high mass sample contains haloes with $\nu > 3.5$. 
The solid lines represent the results from \cite{Diemer2014}. The dotted lines, in the top panels, represent the median density profiles, while the bottom panels show the logarithmic slope profile, both obtained with our model.
The plot shows a good agreement between the profiles predicted by the model and the result of the \citet{Diemer2014} simulated model, only differing by a very small offset and lower numerical noise in our model's case. The very small difference in slope cannot be perceived in the density plots. 
The low-$\nu$ sample median profile's slope changes slowly till $r \gsim \rvir$. A large scatter is visible at larger radii.  The high-$\nu$ sample presents a steep profile at $r \gtrsim 0.5 \rvir$. The slope varies from $-2$ to $-4$ over a radius range restricted to $\approx 4$ rescaled radius, as is shown in the bottom panel representing the slope profiles. 
The plot also shows that an NFW, and an Einasto profiles give a good description of 
the low-$\nu$ sample out to $r \approx \rvir$, while the fast steepening of the slope of the high-$\nu$ sample is poorly described. The NFW and Einasto profiles present radically different behaviors for the outer density profiles of haloes. They are able to fit the low-$\nu$ out to 
$r \approx \rvir$ for the high-$\nu$ haloes, and $r \approx 0.5 \rvir$ in the case of the high-$\nu$ haloes. In order to fit the logarithmic slope profile, it is necessary to use a different fitting formula which was presented by \citep[Sec.~3.3, and in their Appendix]{Diemer2014}. Both in the case of the low-$\nu$ and high-$\nu$ samples, the profiles flatten to a slope of $\approx -1$ at $r \gtrsim 2 \rvir$.

The previous plots in Fig.~\ref{fig:comparison1} shows the profiles of a $\nu$ bin at $z=0$. In general one could expect that the profiles of a given $\nu$ when density and radii are rescaled in the correct way, are self-similar in shape. The problem is to find what are the radii and density to be used for the rescaling.

Fig.~\ref{fig:comparison2} shows the self-similarity of the redshift evolution of the profiles both for the \cite[solid lines]{Diemer2014} and for our model's (dotted lines). The left panel displays the redshift evolution of the median density profiles for a peak with $\nu >3.5$, as a function of the radius rescaled by $R_{200m}$, with density rescaled by $\rho_m$. The central panel represents the same density profiles 
rescaled by $R_{\rm vir}$, and $\rho_{\rm vir}$, respectively. The right panel represents the same density profiles rescaled by $R_{\rm c}$, and $\rho_{\rm c}$, respectively. The black lines corresponds to $z=0$, and the red lines to $z=6$.

We emphasize that the figure plots profiles in physical units, rescaled, in the three panel going from left to right, by $\rtoc$, $\rvir$, and $\rtom$, and the densities rescaled by the corresponding quantities $\rhoc$, $\rho_{\rm vir}$, and $\rho_{\rm m}$. The plot shows clearly that the halos structure is approximately self similar after 
rescaling with $R_{\Delta}$. It is also evident that the self-similarity depends on the kind of rescaling chosen. The most self-similar inner structure of haloes are obtained by rescaling radii and densities with $\rhoc$ and $\rtoc$. Conversely, the most-self-similarity of the outer profiles is obtained rescaling with $\rtom$ and $\rhom$. In order to propose a more readable figure, we only plotted the haloes at two redshift, $z=0$, and $z=6$. We compared the result of our model (dotted line) in Fig.~\ref{fig:comparison3} with \citep{Diemer2014}, and we found again that both sets of results are in agreement, confirming the discussed self-similarity. In order to have a better understanding of the self-similarity, we may use the logarithmic slope profiles. They clearly reveal the radii at which rapid changes in slopes happen. 

Fig.~\ref{fig:comparison3} plots the logarithmic slope of three $\nu$ bins at different redshifts, 
rescaled by $\rtom$ (top row) and $\rtoc$ (bottom row).
The left panels of Fig.~\ref{fig:comparison3} present results for the $\nu >3.5$ sample of Fig.~\ref{fig:comparison2}. The central, and right panels refer to the $\nu$ bins shown in \citep[Fig.~5]{Diemer2014}. Radii in the top panels are rescaled in units of $R_{\rm 200m}$, while those in the bottom panel are expressed in units of $R_{\rm 200c}$. The dotted lines represent the result of our model. The solid lines stand for the \cite{Diemer2014} simulations. The dashed lines provide fits using the Einasto profile. Black, and red lines correspond to different redshifts indicated in the legends of the top panels. In all cases, the slopes show a sharp steepening, followed by a flattening.
%
%
%
In units of $\rtom$, such sharp variations of the slope occur at the same radii, with almost no sign of evolution of the transition. The radius of the steepest slope occurs around $\approx 1-1.2\rtom$ for all $\nu$ and redshifts. Furthermore, for haloes rescaled in units of $\rtom$, the outer flattening displays almost no evolution or variation with $\nu$.
The situation is different in the case $r < \rtom$. In this case, 
a variation with $\nu$ and $z$, of the slopes of the profiles at a given $r / \rtom$, is observed. If radii and densities are rescaled by $\rho_c$ and $\rtoc$, the opposite is valid. The shapes of low-$\nu$ and high-$\nu$ profiles are different, but in any case they show a certain degree of uniformity 
at $r > \rtom$ when rescaled by $\rtom$, and at $r < \rtoc$ when rescaled by $\rtoc$.
The previous discussion brings to the conclusion that the inner profiles are self-similar in units of $r / \rtoc$, while the outer profiles are self-similar in units of $r / \rtom$. It is interesting to note that one would expect that a concentration of haloes should be 
more universal in terms of $\nu$, if the radius definition is related to the critical density. 
As in the previous figure, we only plotted two redshift dependence, so as to obtain a more readable plot. Again the dotted lines represent the result of our model, which is in agreement with that of \citep{Diemer2014}.   
Another issue studied by \citep{Diemer2014} is the dependence on the mass accretion rate. As mentioned, our main goal was to show that the predictions of the \citep{Diemer2014} simulated model are in agreement with our model, namely that our model is able to catch the main characteristics of the density profiles, and their logarithmic slope, correctly describing the behavior of the outer region of the density profiles. For this reason, we do not discuss the dependence on the mass accretion rate. The comparisons shown are sufficient to establish that our model is adequate to describe the outer region of the density profile. 

\section{Conclusions}\label{sec:conclusions}

DM haloes outskirts are characterized by very different density profiles than the inner density profile, usually fitted with models such as the NFW or the Einasto profile. The outer density profile is very steep over a small radial range. This kind of behavior has been interpreted as due to pile up of the orbits of particles, a splashback of material located near half an orbit after collapse. Modeling spherical haloes, such radius provides a sharp separation of infalling matter from material just orbiting the halo, including its satellites. For exact spherical symmetry, a caustic, in this case an infinitely sharp density drop, characterises the splashback radius. In realistic halos, that caustic is smoothed out. The complexity of orbits in $\Lambda$CDM halos, their non-sphericity and sub-structures, make the splashback radius identification in each halo non-trivial. However, it has been observed in weak lensing as well as around stacked clusters in satellite galaxies density profiles.

Theoretically, the location of the splashback radius was obtained through a very simple model
by \citep{Adhikari2014}, calculating the secondary infall of DM shell trajectories within a growing, NFW profiled, DM halo. 
 Since the halo profile was imposed {\it a priory}, instead of calculating it from trajectories of DM shells they were not able to find the DM profile around the splashback radius. 
A more complete model was proposed by \citet{Shi2016}, who extended the \citet{Fillmore1984} self-similar collapse model to a $\Lambda$CDM universe, allowing to calculate simultaneously the trajectories and the DM halo profile. \citet{Diemer2014} studied through simulations the density profiles of $\Lambda$CDM haloes, focusing on the outer regions of the halo. They found a noteworthy deviations of the outer profile from the classical distributions, such as the NFW and Einasto profiles, in the form of a sharper steepening than predicted of the density profile's logarithmic slope. They also found that the outermost density profiles at $r \geq R_{\rm 200 m}$ are self-similar if the radii are rescaled by $R_{\rm 200 m}$. At the same time, the inner density profiles are most self-similar if the radii are rescaled by $R_{\rm 200 c}$. 

In the present paper, we proposed an improved spherical infall model that, differently from 
\citep{Adhikari2014}'s model, takes into account shell crossing, and several other effects, like    
ordered, and random angular momentum, dynamical friction, and adiabatic contraction. Using the model we may obtain the halo density profile, and study the behavior of the profile in its outer regions. \Mov{From Gaussian initial conditions, we generated populations of haloes that can be statistically confronted with other models or observations. }A comparison of the profile and the logarithmic slope profile with the result of the \citet{Diemer2014} simulations shows a good agreement. We then reobtain\Mov{ed} the results \Mov{from }
\citet{Diemer2014} concerning the characteristics of density profile and its logarithmic slope.

\section*{Acknowledgments}
MLeD acknowledges the financial support by the Lanzhou University starting
fund, the Fundamental Research Funds for the Central Universities
(Grant No. lzujbky-2019-25), National Science Foundation of China (grant No. 12047501) and the 111 Project under Grant No. B20063. 
\bibliographystyle{apsrev4-1}
\bibliography{old_MasterBib,}

\end{document}